\documentclass[conference]{IEEEtran}
\IEEEoverridecommandlockouts
\usepackage{cite}
\usepackage{amsmath,amssymb,amsfonts}
\usepackage{algorithmic}
\usepackage{graphicx}
\usepackage{textcomp}
\usepackage{xcolor}
\def\BibTeX{{\rm B\kern-.05em{\sc i\kern-.025em b}\kern-.08em
    T\kern-.1667em\lower.7ex\hbox{E}\kern-.125emX}}

\DeclareMathOperator{\mean}{mean}
\begin{document}

\title{BYOL for Audio: Self-Supervised Learning for General-Purpose Audio Representation\\
}

\author{\IEEEauthorblockN{Daisuke Niizumi, Daiki Takeuchi, Yasunori Ohishi, Noboru Harada, and Kunio Kashino}
\IEEEauthorblockA{\textit{NTT Corporation}, Japan \\
daisuke.niizumi.dt@hco.ntt.co.jp}
}

\maketitle

\begin{abstract}
Inspired by the recent progress in self-supervised learning for computer vision that generates supervision using data augmentations, we explore a new general-purpose audio representation learning approach.
We propose learning general-purpose audio representation from a single audio segment without expecting relationships between different time segments of audio samples.
To implement this principle, we introduce Bootstrap Your Own Latent (BYOL) for Audio (BYOL-A, pronounced "viola"), an audio self-supervised learning method based on BYOL for learning general-purpose audio representation.
Unlike most previous audio self-supervised learning methods that rely on agreement of vicinity audio segments or disagreement of remote ones, BYOL-A creates contrasts in an augmented audio segment pair derived from a single audio segment. With a combination of normalization and augmentation techniques, BYOL-A achieves state-of-the-art results in various downstream tasks. 
Extensive ablation studies also clarified the contribution of each component and their combinations.
\end{abstract}


\section{Introduction}

The recent progress in unsupervised learning in natural language processing and computer vision domains has had a significant impact\cite{brown2020gpt-3}\cite{dosovitskiy2020image}, showing a substantial possibility of exploiting massive data without labels. For these successes, self-supervised learning methods that generate pseudo labels as supervision have played a central role\cite{liu2020selfsupervised}.
In the computer vision domain, contrastive learning, which leverages the instance discrimination pretext task, has become dominant in self-supervised learning\cite{Le-Khac20CRLreview}. It achieves competitive performance compared to conventional supervised learning and even outperforms it in some downstream tasks such as object detection\cite{he2020momentum}\cite{chen2020improved}.

In the contrastive learning setting, training is driven by comparison among positive and negative samples. The positive samples are augmented copies, or views, from the same input, and the negative samples are augmented views from different inputs. In the training process, representations of positive samples in embedding space are mapped closer together, whereas those of positive samples and negative samples are pushed away. However, for achieving better performance, this contrastive learning requires a large number of negative samples to compare\cite{liu2020selfsupervised}. To mitigate this problem, SimCLR\cite{chen20simclr} uses a significant number of batch samples, and MoCo\cite{chen2020improved} operates a large queue to accommodate a larger number of negative samples.

\begin{figure}[tb]
  \centering
  \includegraphics[width=0.95\columnwidth]{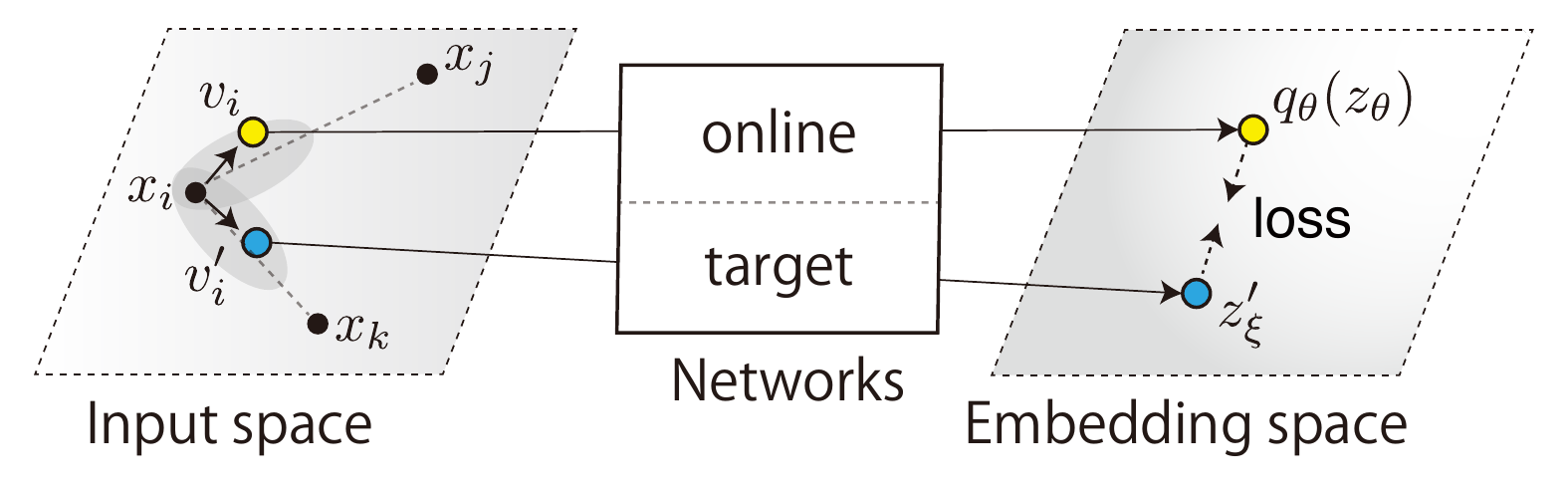} 
      \caption{BYOL\cite{grill2020byol} for audio representation learning scenario.
      A single input audio $x_i$ branches in two directions, or views, $v_i$ and $v_i'$ by mixing audio $x_j$ and $x_k$ and making pitch/time modifications.
      $v_i$, $v_i'$ are projected through networks, and then loss is minimized on the projected embeddings. BYOL updates its online network weights with calculated loss, while it updates target network weights as an exponential moving average of the online counterpart.}
  \label{fig:scenario}
  \vspace{-10pt}
\end{figure}

On the other hand, Bootstrap Your Own Latent (BYOL)\cite{grill2020byol} takes a different approach in that it does not use negative samples. Instead, it directly minimizes the mean squared error of embeddings originating from the same input with contrasts created by data augmentations. This representation learning setting may lead to collapsed representations\cite{liu2020selfsupervised}, but the system architecture and training algorithm can avoid this problem. On the contrary, it claims state-of-the-art (SOTA) performance.

In the audio domain, various self-supervised audio representation learning methods have been proposed\cite{oord2019cpc}\cite{ravanelli2020paceplus}\cite{fonseca2020unsupervised}\cite{jansen2017triplet}\cite{shor2020trill}. In particular, COLA\cite{saeed2020cola} learns general-purpose representations and outperforms previous methods. Many of these methods utilize the time-series aspect of audio signals: audio segments cropped closer in time are expected to have closer representations, whereas those far away in time are expected to have distanced representations.
This is conceived to be a rational expectation, but contradictory use cases can be found easily.
For example, repetitive sounds like music could have similar contents in the remote time segments because music compositions, by their nature, repeat motifs.
On the other hand, 
short acoustic events (e.g., a single knock, a gunshot) can occur in a short duration; thus, even adjacent segments (e.g., a knock followed by a footstep) can make differences in contents for acoustic events.

We think these are the fundamental problems caused by expecting relationships among multiple segments. In addition, similar problems can also happen when we use contrastive learning\cite{fonseca2020unsupervised}\cite{saeed2020cola} or triplet loss\cite{jansen2017triplet}\cite{shor2020trill} because the comparison of multiple samples is the core of their loss calculation.

%

We address these problems by having general-purpose audio representations learned from a single audio segment instead of from a comparison of multiple segments.
In addition, the use of contrastive or triplet loss has to be avoided. This consequently requires the use of BYOL with effective audio data augmentations.
For augmentations, we focus on learning 1) the foreground acoustic event sound (e.g., dog barking, gunshot) as a dominant sound representation, and 2) the sound texture details for describing general-purpose representation.

\begin{itemize}
    \item The better foreground sound representation is supposed to be consistent regardless of the background sound contents. Then it can be better learned from samples with random background variations while the foreground is kept unchanged.
    Considering a natural sound is a mixture of sounds from various sources, mixing small amount of sounds can approximate making variations on the background. Therefore, we adopt mixup\cite{zhang2018mixup}, which mixes samples within a dataset.
    \item Sounds from an acoustic scene or a sound texture can vary their pitch/speed/time, while the details can be consistent.
    This suggests that the details can be learned under the random variations of pitch/speed/time shifts; thus, we use approximation of audio pitch shifting and time stretching techniques\cite{timepitch} for this purpose. We expect learned representation to compress useful information of sound details in order to serve various general tasks.
\end{itemize}
These techniques were used in a similar way in \cite{fonseca2020unsupervised}, but sources from different time segments were cropped. We have clear purpose regarding what information is to be learned, and we create changes on a pair of segments originating from exactly the same segment, not from multiple segments.\\



\vspace{-5pt}
The contributions of this paper are as follows:

\begin{itemize}
    \item We propose learning general-purpose audio representations from a single audio segment without expecting relationships between different time segments of audio samples.
    \item We propose a self-supervised learning method named BYOL for Audio (BYOL-A, pronounced "viola"). The method learns representations from a single audio segment input with a dedicated audio augmentation module that focuses on foreground and content details.
    It outperforms previous methods that learn from contrast of segments derived from different times.
    \item We propose to learn foreground sound by combining pre-normalization and mixup, while learning content details through approximation of pitch shifting and time stretching. An extra post-normalization is also applied to compensate for statistical drift caused by augmentations.
    \item We conduct extensive ablation studies that make clear the contributions of each block in the BYOL-A augmentation module.
\end{itemize}

Fig. \ref{fig:scenario} illustrates the entire scenario of self-supervised learning from a single segment input.
An input audio segment is once normalized. Then, two augmented copies, or views, are created by adding another sample as background sound and modifying pitch and time. These views are processed through parallel networks (i.e., online and target networks). Then, the online network is updated with loss calculated from projected embeddings, and the target network is updated as an exponential moving average of the online network. The system gradually learns to produce better representations by repeating these training processes.

\section{Related Work}

\subsection{Self-supervised learning schemes}

In the image domain, self-supervised learning variants that make use of data augmentation techniques have been proposed. Contrastive learning methods such as SimCLR\cite{chen20simclr} 
use a big batch size, while MoCo\cite{he2020momentum}
\cite{chen2020improved} has a FIFO queue to accommodate negative samples. Both claim SOTA performance. On the other hand, BYOL\cite{grill2020byol} also claims SOTA performance, though they use no negative samples.
Among these methods, BYOL meets our needs for learning from a single input without the use of contrastive loss.

Methods that combine self-supervised learning and mixup have also been proposed. Domain-agnostic contrastive learning (DACL)\cite{verma20dacl} proposes a mixup variant \textit{Mixup-noise} for contrastive learning setting.
i-MIX\cite{lee2020imix} is a contrastive learning method that follows more of the original concept of applying mixup to both the features and its loss calculation.
Fonseca et al. \cite{fonseca2020unsupervised} proposed a contrastive learning approach for sound event representations and adopted a mixup variant \textit{mix-back} to add background noise to input log-mel spectrogram audio.
All these methods are based on contrastive loss that compares positive and negative samples, and this is a fundamental difference from our approach. Experiments on i-MIX have also been conducted using BYOL and audio, as one of the multimodal applications of its domain-agnostic approach. However, audio has not been tested extensively, and the basic concepts are different from the present work; for instance, its usage of mixup is not concerned with audio content.

In the audio domain, many methods rely on the relationships between segments cropped from different times.
CPC\cite{oord2019cpc} uses contrastive learning with future representations predicted from a series of past segments.
Jansen et al. \cite{jansen2017triplet} uses triplet loss for learning with augmentation techniques, including adding random noise, time/frequency translation, example mixing, and temporal proximity.
TRILL\cite{shor2020trill} learns representation by using triplet loss based on \cite{jansen2017triplet} so that segments that are closer in time are also closer in the embedding space.
COLA\cite{saeed2020cola} is a contrastive learning method for general-purpose audio representation that handles segments from the same clip as positives and segments from others as negatives.
Among conventional methods for downstream tasks we tested, TRILL and COLA show SOTA performance.

Cross-modal methods that input audio and other modalities are also related.
COALA\cite{favory2020coala} uses tags (labels) accompanied with audio and uses contrastive loss to maximize co-alignment of both.
$L^3$-Net\cite{cramer2019openl3} (or OpenL3) inputs video along with audio and learns representations from the correspondence between the audio and video.
These approaches show reference performance when non-audio data is used.

\begin{figure*}[tb!]
  \centering
  \includegraphics[width=0.9\textwidth]{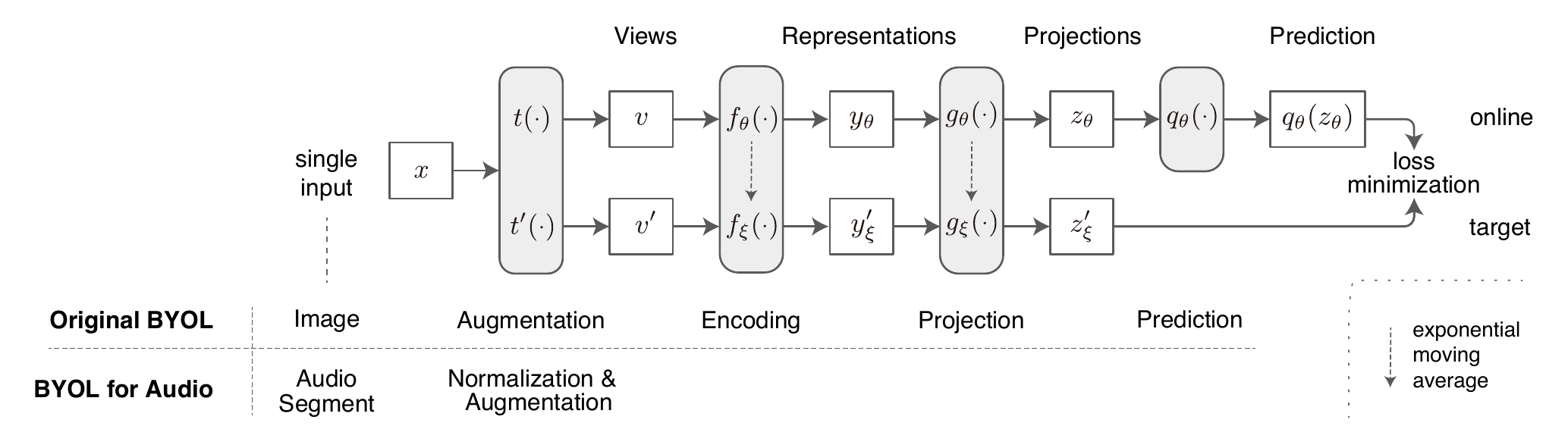} 
  \vspace{-7pt}
  \caption{BYOL and BYOL-A system overview.}
  \label{fig:system}
  \vspace{-7pt}
\end{figure*}

\begin{figure}[tb!]
  \centering
  \includegraphics[width=0.9\columnwidth]{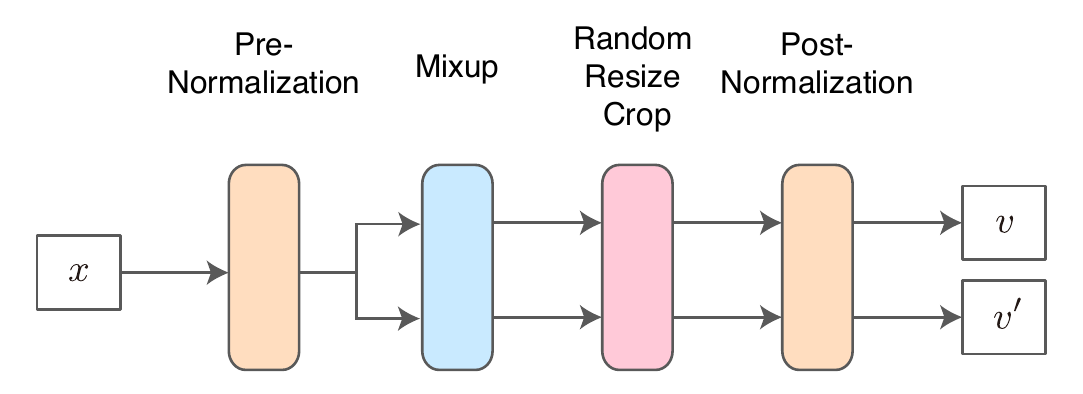} 
  \vspace{-7pt}
  \caption{Audio augmentation module of BYOL-A.}
  \label{fig:aug}
\end{figure}

\begin{figure}[tb!]
  \centering
  \includegraphics[width=0.6\columnwidth]{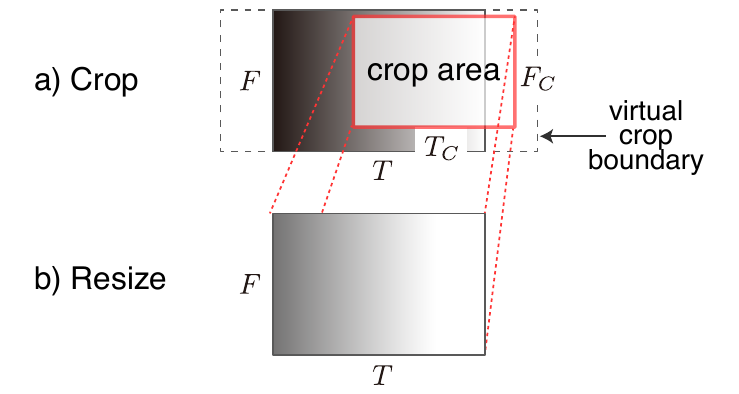} 
  \vspace{-5pt}
  \caption{Random resize crop of spectrogram. a) Randomly chosen crop area content is b) resized to the same size as the input.}
  \label{fig:rrc}
  \vspace{-7pt}
\end{figure}





\subsection{Bootstrap Your Own Latent (BYOL)}

BYOL is an algorithm for self-supervised learning of image representations.
As shown in Fig.~\ref{fig:system}, it consists of two neural networks, referred to as online and target networks.
The online network is defined by a set of weights $\theta$, and the target network has the same architecture as the online network but uses a different set of weights $\xi$.
First, BYOL produces two augmented views, $v\triangleq t(x)$ and $v'\triangleq t'(x)$, from an image $x$ by applying respectively image augmentations $t\sim \mathcal{T}$ and $t'\sim \mathcal{T}'$, where $\mathcal{T}$ and $\mathcal{T}'$ denote the two distributions of the image augmentations.
Then, the online network outputs a representation $y_{\theta}$, a projection $z_{\theta}$, and a prediction $q_{\theta}(z_{\theta})$  from the first view $v$.
On the other hand, the target network outputs $y'_{\xi}$ and the target projection $z'_{\xi}$ from the second view $v'$.
Finally, the following mean squared error between the L2-normalized predictions $\overline{q_\theta}(z_\theta)$ and target projections $\overline{z}'_\xi$ is calculated:

\vspace{-10pt}
\begin{equation}
L_{\theta,\xi} \triangleq ||\overline{q_\theta}(z_\theta) - \overline{z}'_\xi||^2_2 = 2 - 2 \cdot \frac{\langle q_\theta(z_\theta), z'_\xi \rangle }{||q_\theta(z_\theta)||_2 \cdot ||z'_\xi||_2},
\label{eq:eq-byol-mse}
\end{equation}
where $\langle\cdot, \cdot\rangle$ denotes the inner product.
To symmetrize the loss $L_{\theta,\xi}$, $L'_{\theta,\xi}$ is computed by feeding $v'$ to the online network and $v$ to the target network. The final loss is defined as $L_{\theta,\xi}^{\mathrm{BYOL}} = L_{\theta,\xi} + L'_{\theta,\xi}$.
At each training step, BYOL minimizes this loss function with respect to $\theta$ only, but $\xi$ is a slowly moving exponential average of $\theta$: $\xi \leftarrow \tau \xi + (1 - \tau) \theta$, where $\tau$ is a target decay rate.

It has been empirically shown that the combination of adding the predictor to the online network and using the moving average of the online network parameters as the target network encourages encoding more and more information within the online projection and avoids collapsed solutions such as constant representations.

\section{BYOL for Audio (BYOL-A)}\label{sec:byola}

We propose learning general-purpose audio representations from a single audio segment without expecting relationships between different time segments of audio samples. To implement this principle, we introduce BYOL-A.

As shown in  Fig.~\ref{fig:system}, we extend BYOL\cite{grill2020byol} for general-purpose audio representation learning.
In BYOL-A, we input audio preprocessed as a log-scaled mel-spectrogram, a time-frequency feature, because a typical encoder convolutional neural network accepts time-frequency features and converts them into representation embeddings. 
In addition, we replace the augmentation module in BYOL with ours so that the learning system can handle audio and create contrasts in augmented views for learning general-purpose audio representations.

As shown in Fig.~\ref{fig:aug}, the BYOL-A augmentation module consists of four blocks.
First, the Pre-Normalization block normalizes a single input audio segment so that following augmentation processes, especially mixup, become stable.
The normalized input is duplicated into two copies and fed to the following Mixup block.
Then the Mixup block creates two outputs that are mixes of normalized inputs and randomly chosen past normalized inputs.
The following Random Resize Crop (RRC) block resizes and crops the outputs randomly, and finally the Post-Normalization block adjusts statistical drifts caused by the former augmentations.

For learning general-purpose audio representations, we focus on foreground acoustic events and all content details.
The Mixup block is designed to create contrast for learning foreground acoustic event representations, and it is combined with the Pre-Normalization block for stable performance gain. The RRC block approximates pitch shifting and time stretching in time-frequency features for learning generic representations of content details.


\subsection{Pre-Normalization}
Input data $x$ is normalized to $\Tilde{x} = \frac{x - \mu}{\sigma}$, where $\mu$ and $\sigma$ are the average and standard deviation of training samples respectively.

This normalization stabilizes computations in the system in two ways.
One is by mitigating augmentation parameter sensitivity, which enables following blocks to assume that input range virtually follows $N(0, 1)$. The other is by normalizing statistical differences between training datasets.

\subsection{Mixup for foreground acoustic event}

Using normalized log-mel spectrogram audio as input, the Mixup block mixes past randomly selected input audio in a small ratio. As a result, added audio becomes a part of the background sound in the mixed audio. This produces contrast in the background sound in the pair of Mixup outputs, which in turn encourages learning representations of invariant foreground acoustic event sounds.
This is similar to \textit{mix-back}\cite{fonseca2020unsupervised}, which adds a random sample from a dataset as background sound, but the purpose of the \textit{mix-back} is to create a set of positive samples sharing less information in the contrastive learning setting.

We use the basic mixup calculation as an augmentation technique.
While mixup was originally designed for mixing both features and labels, we apply it to audio features only (because of the absence of labels). In addition, as audio is log-scaled, we convert input to a linear scale before the mixup calculation and convert it back to a log-scale again. In this paper, we refer to these operations as log-mixup-exp, from the analogy to the log-sum-exp\cite{logsumexp} calculation. Log-mixup-exp of $i$th input $x_i$ is

\vspace{-10pt}
\begin{equation}
\Tilde{x}_i = \log{\left((1 - \lambda) \exp(x_i) + \lambda \exp(x_k)\right)}
\label{eq:eq-log-mixup}
\end{equation}

where $x_k$ is a mixing counterpart, and mixing ratio $\lambda$ is sampled from uniform distribution $U(0.0, \alpha)$ instead of from a beta distribution in the original mixup.
In addition, $\alpha$ is a mixing ratio hyper-parameter that controls the degree of contrast between the resulting two mixed outputs. We observed that the evaluation result improves with smaller $\alpha$, $0.4$ for example, where $\Tilde{x}_i$ keeps the original contents $x_i$ more than its counterpart $x_k$, as we found in preliminary experiments.

$x_k$ is randomly chosen from a memory bank, a FIFO queue, storing past inputs. As input is randomly sampled from the training dataset, queued samples in the memory bank form random subsets of the dataset. We store $2,048$ samples in the memory bank, which is larger than the batch size and big enough to maintain randomness.



\subsection{RRC for all content details}

RRC is an image augmentation technique we use as an approximation of pitch shifting and time stretching of input audio log-mel spectrograms for learning a representation of all content details.
We expect details spread over a spectrogram to be learned regardless of the created differences in pitch and time among outputs.


Fig.~\ref{fig:rrc} shows the random crop procedure. The unit size of the input spectrogram consists of a number of frequency bins, $F$, and a number of time frames, $T$. First, we sample the random crop area from the virtual crop boundary, which has longer time frames than the input, $1.5 \times T$ for example. The size of the crop area is randomly sampled as
$$F_C = \lfloor\min(U(h_1, h_2), 1.0) \times F\rfloor$$
$$T_C = \lfloor U(w_1, w_2) \times T\rfloor$$
where $F_C$ and $T_C$ are the number of frequency bins and number of time frames of random crop size, respectively, $h_1$ and $h_2$ form a frequency bin range $[h_1, h_2]$, $w_1$ and $w_2$ form a time frame range $[w_1, w_2]$, $\lfloor\cdot\rfloor$ is a floor function, and $\min(\cdot, \cdot)$ is a minimum function.
Contents in the crop area are then resized to the size of the input with bicubic interpolation. The virtual crop boundary is wider than the input and we use $[0.6, 1.5]$ for both frequency bin and time frame ranges in this paper, so the crop area can contain the outside of the input. This area is filled with zeros.
Note that we do not crop the outside of the frequency bin, which is restricted by the $\min()$ function in the $F_C$ calculation above.


\subsection{Post-Normalization for statistical drift adjustment}

Augmentation operations before this block can cause statistical drift in their outputs. This block adjusts the drift so that final output views of the BYOL-A augmentation module become $\sim N(0, 1)$.
The calculation is done in the same manner as pre-normalization, but uses average and standard deviations calculated from batch samples.

\section{Experiments}
We assessed the representations learned by BYOL-A by conducting experiments with six audio downstream tasks under a linear evaluation protocol. The downstream tasks have both audio samples and labels, and the evaluation result is the accuracy of a linear model trained in a supervised setting with representation feature embeddings as input. These feature embeddings were converted from audio samples using a frozen encoder network pretrained by BYOL-A.


In all experiments, BYOL-A was pretrained on AudioSet \cite{gemmeke2017audioset}, a large scale dataset commonly used in previous studies.

\subsection{Experimental Setup}

We repeated the cycle of pretraining and evaluation and averaged the results. The number of cycles was three for pretraining on the full AudioSet;
it was five for pretraining on a 1/10 subset of AudioSet or FSD50K\cite{fonseca2020fsd50k}.

\subsubsection{Audio data format and conversion parameters}
We converted all sound clips to a log-scaled mel spectrogram with a sampling frequency of 16,000 Hz, window size of 64 ms, hop size of 10 ms, and mel-spaced frequency bins $F=64$ in the range 60–7,800 Hz.
The number of frames, $T$, in one segment was 96 in pretraining, which corresponds to 1,014 ms. A segment of shape $F \times T$ was randomly cropped from each audio clip and used in pretraining.
For the downstream tasks, the number of frames, $T$, in one segment was determined by the average duration of each dataset (e.g., 400 frames for NSynth with average duration of 4.0 s.).
A segment of shape $F \times T$ was randomly cropped from each audio clip and encoded for linear evaluation in the downstream tasks. Shorter clips were padded with zeros at both the head and tail.

\subsubsection{Encoder network}
We used a simple CNN based on a network used in a solution of Task 6 (Automated Audio Captioning)
of the Detection and Classification of Acoustic Scenes and Events (DCASE) 2020 Challenge\cite{koizumi2020t6ntt} \cite{takeuchi2020effects}.
Though this network is simpler and smaller than networks from the image domain (e.g., ResNet\cite{he2015resnet}, EfficientNet-B0\cite{tan2020efficientnet}) used in previous 
studies \cite{shor2020trill} \cite{saeed2020cola},
we think this is a realistic design choice because smaller networks have been practically used in audio machine learning studies \cite{koizumi2020t6ntt}\cite{kameoka2020voicegrad}\cite{zhang2018deep}.
In addition, the experimental results in this paper show that the performance of our CNN is good enough to score SOTA.

The CNN has a variable hyper-parameter, namely the dimension size of representation embedding, that is also used as the size of the linear layers. We varied this hyper-parameter in the experiments to see the performance changes.

The details of the network are described in appendix \ref{appendix:encoder}.

\subsubsection{BYOL-A pretraining details}

Projection and prediction in BYOL-A networks are the same multilayer perceptrons (MLPs) in the original BYOL, i.e., a linear layer with output size of 4,096 followed by batch normalization \cite{ioffe15batch}, rectified linear units (ReLU), and a linear layer to output embeddings with 256 dimensions.
We used the Adam optimizer with a learning rate of $0.0003$, target decay rate parameter $\tau = 0.99$, and batch size of $256$, and it was trained for 100 epochs.

For augmentation blocks, we used mixup $\alpha=0.4$. The size of the virtual crop boundary has the number of time frames to $1.5$ times the input and the number of frequency bins to the same as the input. The crop range is $[0.6, 1.5]$ for both frequency bins and time frames.
All these values were found in a preliminary parameter search using Optuna\cite{akiba2019optuna}. We used a single set of augmentation $\mathcal{T}$, i.e., $t, t' \sim \mathcal{T}$.

The number of pretraining samples collected from balanced\_train\_segments and unbalanced\_train\_segments data splits in the AudioSet\cite{gemmeke2017audioset} dataset was 1,963,807 in total. No labels were used in pretraining.
In the ablation studies, a $1/10$ subset of AudioSet (210,315 samples in total) was used in pretraining.

\subsubsection{Downstream tasks}
Below are the tested tasks. These tasks were also used in previous studies\cite{shor2020trill}\cite{saeed2020cola}\cite{favory2020coala}\cite{cramer2019openl3}, so evaluation results can be compared.

\begin{itemize}
\item NSynth (NS) \cite{nsynth2017} dataset as musical instrument family classification with 11 family name classes for 305,979 samples and average duration of 4.0 s.
\item UrbanSound8K (US8K)\cite{salamon2014urbansound} dataset as sound classification with ten acoustic scene classes for 8,732 samples and average duration of 4.0 s. This task dataset has predefined splits with ten folds. We followed leave-one-out cross validation of these ten folds to get the average accuracy.
\item VoxCeleb1 (VC1)\cite{voxceleb} dataset as speaker identification task with 1,211 speaker ID classes for 153,514 samples and average duration of 8.2 s.
\item VoxForge (VF)\cite{voxforge} dataset as language identification task with six language ID classes for 176,438 samples and average duration of 5.8 s.
\item Speech Commands V2 (SPCV2)\cite{speechcommandsv2} dataset as command classification task with 35 command word classes for 105,829 examples and duration of 1.0 s.
\item The same Speech Commands V2 dataset, but with 12 command word classes (SPCV2/12). Unlike SPCV2, classes consist of ten basic word selected from 35 words, as well as new classes \textit{silence} and \textit{others}. The \textit{silence} audio samples were created from background noise sounds, which are not used as a class in SPCV2, and \textit{others} contains all other 25 class samples from SPCV2. This setup results in a highly imbalanced number of class samples and additional complexity compared to SPCV2 in spite of the smaller number of classes.
\end{itemize}

\subsubsection{Linear evaluation details}
In the linear evaluation protocol, a single linear layer is trained to fit the downstream dataset.
We trained the linear model using Adam optimizer with a learning rate of $0.001$, for a $200$ epoch maximum with early stopping of ten epochs. We ran each evaluation ten times and averaged the accuracy results.

\subsubsection{Comparison with COLA\cite{saeed2020cola}}

COLA was specifically reproduced with our implementation, denoted as COLA'. COLA takes the opposite approach from ours: It maximizes agreement of two random cropped segments of the same sound clip, uses contrastive loss for comparison with negative samples, and includes no data augmentations.

COLA' has an extra normalization module, the same as BYOL-A, and the same encoder network as BYOL-A. This enables us to compare results between a single-segment input and two-segment input and evaluate the effectiveness of BYOL-A augmentation blocks under the two-segment input setup of COLA'.
We used a version of COLA with bilinear similarity\cite{saeed2020cola} that has better performance. Pretraining parameters were the same as with BYOL-A pretraining, except batch size was $1,024$, four times larger than with BYOL-A, for ensuring better performance under contrastive learning.

\begin{table*}[htb!]
\vspace{-10pt}
\caption{Performance comparison results for downstream tasks}
\vspace{-5pt}
\label{tab:result-dstasks}
\centering
\begin{tabular}{rrr|cccccc|c}
\hline
Method & Dim. & Remarks & NS &  US8K & VC1 & VF & SPCV2/12 & SPCV2 & Average \\
\hline
\hline
TRILL\cite{shor2020trill}  &  & conventional &   N/A &   N/A &    17.9\% &    88.1\% & 74.9\% &   N/A &     N/A \\
COLA\cite{saeed2020cola}  & & conventional & 63.4\% &   N/A &    29.9\% &    71.3\% & 71.7\% & 62.4\% &     N/A \\
OpenL3 \cite{cramer2019openl3} $^{\mathrm{1}}$  & & reference &    N/A & 78.2\% &      N/A &      N/A &   N/A &   N/A &     N/A \\
COALA\cite{favory2020coala} $^{\mathrm{2}}$  & & reference &  73.1\% & 72.7\% &      N/A &      N/A &   N/A &   N/A &     N/A \\
\hline
COLA'  & 512-d  & our impl. &  65.4\% & 76.3\% &    25.9\% &    73.5\% & 59.1\% & 63.1\% &   60.5\% \\
COLA'  & 1024-d & our impl. &  69.3\% & 77.1\% &    31.2\% &    76.7\% & 71.9\% & 71.0\% &   66.2\% \\
COLA'  & 2048-d & our impl. &  70.2\% & 78.5\% &    30.4\% &    79.5\% & 76.7\% & 76.8\% &   68.7\% \\
\hline
BYOL-A & 512-d  & \textbf{proposed} &  69.1\% & 78.2\% &    33.4\% &    83.5\% & 86.5\% & 88.9\% &   73.3\% \\
BYOL-A & 1024-d & \textbf{proposed} &  72.7\% & 78.2\% &    38.0\% &    88.5\% & 90.1\% & 91.4\% &   76.5\% \\
BYOL-A & 2048-d & \textbf{proposed} &  \textbf{74.1\%} & \textbf{79.1\%} & \textbf{40.1\%} & \textbf{90.2\%} & \textbf{91.0\%} & \textbf{92.2\%} & \textbf{77.8\%} \\
\hline
\multicolumn{9}{l}{$^{\mathrm{1}}$ Reference results pretrained with audio+video and trained with MLP instead of linear layer.}\\
\multicolumn{9}{l}{$^{\mathrm{2}}$ Reference results pretrained with audio+tag and trained with MLP instead of linear layer.}\\
\end{tabular}
\end{table*}

\begin{table*}[htbp]
\vspace{-8pt}
\caption{Ablations of BYOL-A augmentation module with accuracy results, pretrained with 1/10 AudioSet}
\vspace{-5pt}
\label{tab:result-augs}
\centering
\begin{tabular}{r|cccccc|cc}
\hline
Augmentation blocks used & NS &  US8K & VC1 & VF & SPCV2/12 & SPCV2 & Average &  Degradation \\
\hline
\hline
Mixup+RRC (BYOL-A) &   \textbf{71.2\%} & 77.0\% &     31.0\% &     83.1\% &  \textbf{84.5\%} &  87.2\% &    \textbf{72.3\%} &           \\
\hline
Mixup+Gaussian+RRC  &   69.5\% & 74.3\% &     25.2\% &     \textbf{84.0\%} &  82.8\% &  \textbf{87.4\%} &    70.5\% & BYOL-A -1.8 \\
Gaussian+RRC        &   69.7\% & 73.1\% &     29.2\% &     83.1\% &  78.0\% &  83.1\% &    69.3\% & BYOL-A -3.0 \\
RRC                 &   69.4\% & \textbf{77.1\%} &     \textbf{34.5\%} &     80.3\% &  71.4\% &  77.4\% &    68.4\% & BYOL-A -3.9 \\
Mixup               &   55.6\% & 69.4\% &     22.3\% &     78.3\% &  75.8\% &  82.0\% &    63.9\% & BYOL-A -8.4 \\
Gaussian            &   29.5\% & 31.2\% &      0.9\% &     57.9\% &   9.4\% &  10.3\% &    23.2\% & BYOL-A -49.1 \\
\hline
\end{tabular}
\vspace{-5pt}
\end{table*}

\subsection{Experimental results: Comparison with previous methods}

Table \ref{tab:result-dstasks} shows the results for previous methods (TRILL and COLA), reference methods (OpenL3\cite{cramer2019openl3}, pretrained with audio+video, and COALA\cite{favory2020coala}, pretrained with audio+tag), COLA' (our implementation of COLA), and our proposed BYOL-A. 
For BYOL-A and COLA', we varied the dimensions of representation embeddings as 512, 1,024, and 2,048, which also increases encoder network capacity.

As shown in the Table \ref{tab:result-dstasks}, BYOL-A with 2,048 dimensions outperforms other methods in all tasks. It shows the average result of $77.8\%$, which surpasses the $68.7\%$ for COLA' with 2,048 dimensions.
While BYOL-A with 2,048 dimensions exhibits the best results, embeddings with 512 dimensions also shows competitive performance, especially in the speech command tasks.
\subsection{Ablation study: Contribution of data augmentations}\label{sec:ablation}

In this experiment, we tested various combinations of data augmentation blocks with BYOL-A with 512 dimensions, which was pretrained with 1/10 AudioSet.
We kept the Pre- and Post-Normalization blocks, and replaced augmentation blocks between them.
We also used a Gaussian-noise augmentation block that interpolates training input with random data points sampled from the normal distribution. This was done for comparison with mixup that interpolates within dataset samples.
The Gaussian-noise block sampled from $N(0, 0.4)$, the best parameter in a preliminary test, and followed the log-mixup-exp calculation.

Table \ref{tab:result-augs} shows the results for combining augmentations.

\subsubsection{Contribution of mixup compared with Gaussian-noise}

If we focus on the average result for the Gaussian-noise block, it improves $0.9$ from the RRC's $68.4\%$ to Gaussian+RRC's $69.3\%$.
On the other hand, mixup improves $3.9$ from RRC's $68.4\%$ to Mixup+RRC (BYOL-A)'s $72.3\%$.
However, if we add Gaussian-noise on top of Mixup+RRC, Mixup+Gaussian+RRC degrades average performance $1.8$ down to $70.5\%$.
This empirically shows that mixup's interpolating with samples within dataset works effectively in the BYOL-A setting, whereas interpolating with random data points is not effective.

\subsubsection{Contribution of mixup, RRC, and their combination}

When only Gaussian noise was used, the average result was $23.2\%$; representations cannot achieve sufficient performance.
With mixup only, it improves to $63.9\%$, especially with a larger performance gain on speech command SPCV2. 
The use of mixup consistently gains performance for SPCV2 tasks in other results also, indicating that mixup is effective for learning foreground sound representation, because the SPCV2 dataset is dominant with clear word utterances.
With RRC only, the average result improves up to $68.4\%$ and shows competitive performance among all tasks, indicating that RRC is effective for learning general-purpose representations.
Finally, the average result for Mixup+RRC (BYOL-A) is $72.3\%$, the best average performance among all combinations. This shows that mixup and RRC work complementarily on average, except for a performance drop in the VC1 task.
The VC1 task is speaker identification from real-world random utterances of 1,211 celebrities; the class label is more related to the speaker's voice characteristics than to the utterance content. The recordings have relatively low background noise. Thus, we think the details of textures like consonant or breathing sounds is more important for the VC1 task. Therefore, just applying RRC was better than the use of both mixup and RRC; mixup can add noise to the details.

\subsection{Ablation study: Contribution of normalization blocks}
\label{appendix:ablation-norm}

\begin{table}[tb!]
\vspace{-10pt}
\caption{Ablations of normalization blocks with average accuracy results, pretrained on 1/10 AudioSet}
\vspace{-5pt}
\label{tab:result-norm}
\centering
\begin{tabular}{l|r|r}
\hline
Method &  Average &  Degradation \\
\hline
\hline
BYOL-A                             &    \textbf{72.3\%} &         \\
\hline
w/o Post-Norm                      &    72.1\% & BYOL-A -0.2 \\
w/o Pre-Norm (mixup $\alpha=0.05$) &    70.5\% & BYOL-A -1.8 \\
w/o Pre-Norm (mixup $\alpha=0.1$)  &    70.3\% & BYOL-A -2.0 \\
w/o Pre-Norm (mixup $\alpha=0.4$)  &    68.9\% & BYOL-A -3.4 \\
\hline
\end{tabular}
\vspace{-10pt}
\end{table}

In this experiment, we assessed the contribution of Pre- and Post-Normalization blocks by removing one of them. We avoided removing both to keep the basic configuration of the machine learning setup.
Table \ref{tab:result-norm} shows that removing pre-normalization degrades performance, ranging from  $-1.8$ to $-3.4$, which is a larger impact than removing post-normalization (degradation of $-0.2$).

The reason for the larger degradation observed with the removal of pre-normalization is related to the log-mixup-exp calculation. In this calculation, the log-mel spectrogram is once expanded to a linear scale by $\exp(\cdot)$, mixed with other samples for random degree $\lambda \sim U(0.0, \alpha)$, and then compressed again with $\log(\cdot)$. Then the effect of mixup and its hyper-parameter $\alpha$ depends on the range of the log-mel spectrogram, because the output of mixup is compressed by following $\log(\cdot)$. Pre-normalization helps to stabilize the effect of log-mixup-exp by making the range of the spectrogram constant.
The mixup $\alpha=0.4$ was the sweetest spot found in the preliminary parameter search, and "w/o Pre-Norm" results show that the sweet spot of mixup $\alpha$ drifts, and it does not recover the best performance of $72.3\%$ even when we set $\alpha$ down to $0.05$.

The post-normalization helps prevent degradation caused by statistical drift caused by augmentations.

In summary, the combination of normalizations and augmentations contributes to both performance gain and recovery.

\section{Conclusion}

In this paper, we proposed a self-supervised learning method called BYOL for Audio (BYOL-A), a version of BYOL extended to the audio domain, which learns representations from a single segment audio, and showed its state-of-the-art performance.
The augmentation module of BYOL-A consists of Normalization, Mixup and Random Resize Crop (RRC) blocks. The mixup is effective for learning representations of foreground acoustic event sounds, while RRC works effectively for general-purpose audio representation, and applying both works complementarily.
The Pre- and Post-Normalization blocks work for performance gain of mixup and the recovery from statistical drift.
As a result, all these modules work together as a whole augmentation module in BYOL-A to surpass the previous state-of-the-art results.


The expectation of agreement or disagreement of multiple audio segments was shown to be effective in previous studies, but in this study, it was found that representation learning from a single segment is possible, and it even outperforms former methods.


\appendix
\def\thesection{\Alph{section}}

In this appendix, we explain the details of experiments and additional experiments with analysis.
In appendix \ref{appendix:encoder}, we explain the details of the encoder network and parameter settings.
In appendix \ref{appendix:experiment-cola}, we assess the effectiveness of the BYOL-A augmentation module by applying it to COLA'.
Finally, in appendix \ref{appendix:fsd50k}, we show the performance of BYOL-A pretrained on the FSD50K dataset and compare results.

\subsection{Details of encoder network} \label{appendix:encoder}

\begin{table}[b!]
\vspace{-15pt}
\caption{Encoder network architecture (2048-d)}
\label{tab:cnn-arch}
\vspace{-5pt}
\centering
\begin{tabular}{rr|r|r}
\hline
             Layer-\#    & Layer prms. &         Output shape    &  Parameters \\
\hline
\hline
            Conv2D-1    & 3x3@64 &     [B, 64, 64, 96]  &        640\\
       BatchNorm2D-2    &     &     [B, 64, 64, 96]     &        128\\
              ReLU-3    &     &     [B, 64, 64, 96]     &          0\\
         MaxPool2D-4    & 2x2,stride=2 & [B, 64, 32, 48] &        0\\
            Conv2D-5    & 3x3@64 &     [B, 64, 32, 48]  &     36,928\\
       BatchNorm2D-6    &     &     [B, 64, 32, 48]     &        128\\
              ReLU-7    &     &     [B, 64, 32, 48]     &          0\\
         MaxPool2D-8    & 2x2,stride=2 & [B, 64, 16, 24] &        0\\
            Conv2D-9    & 3x3@64 &     [B, 64, 16, 24]  &     36,928\\
      BatchNorm2D-10    &     &     [B, 64, 16, 24]     &        128\\
             ReLU-11    &     &     [B, 64, 16, 24]     &          0\\
        MaxPool2D-12    & 2x2,stride=2 & [B, 64, 8, 12]  &        0\\
            Reshape-13  &     &        [B, 12, 512]     &          0\\
           Linear-14    & out=2048 &   [B, 12, 2048]    &   1,050,624\\
             ReLU-15    &     &        [B, 12, 2048]    &           0\\
          Dropout-16    & 0.3 &        [B, 12, 2048]    &           0\\
           Linear-17    & out=2048 &   [B, 12, 2048]    &   4,196,352\\
             ReLU-18    &     &        [B, 12, 2048]    &           0\\
   $\max(\cdot)\oplus\mean(\cdot)$-19    &     &            [B, 2048]    &           0\\
\hline
\end{tabular}
\end{table}

Table \ref{tab:cnn-arch} shows the architecture of the encoder convolutional neural network based on a network used in a solution of Task 6 (Automated Audio Captioning) at DCASE 2020 Challenge\cite{koizumi2020t6ntt} \cite{takeuchi2020effects}, where input shape is $[B, 1, 64, 96]$, $B$ is batch size; a single channel, $64$ frequency bins, and $96$ time frames.

Input is compressed by three sets of convolutional blocks of all $64$ channels with stride of $2$. Then, it is reshaped to $[B, 12, 512]$, where time frames $12 = \frac{96}{2 \cdot 2 \cdot 2}$ and feature dimensions $512 = 64  \cdot \frac{64}{2 \cdot 2 \cdot 2}$. The following linear layers upscale the size of feature dimensions to the  hyper-parameter $d$, 2,048 dimensions for example. Then, in the last layer, final output embedding $y$ is calculated as $y = \max(x, 1) \oplus \mean(x, 1)$, where $x$ is input to this calculation, $\max(x, 1)$ is max operation along the time axis, $\mean(x, 1)$ is averaging along the time axis, and $\oplus$ is an element-wise sum.

This network outputs representation embeddings with a fixed shape $(d,)$.
The total numbers of network parameters are 600,192 (512-d), 1,649,792 (1024-d) and 5,321,856 (2048-d).

\subsection{Experiments on BYOL-A augmentation blocks with COLA'+}
\label{appendix:experiment-cola}

\begin{table}[tb!]
\vspace{-5pt}
\caption{Average accuracy results of augmentation blocks on COLA'+ and BYOL-A, pretrained with 1/10 AudioSet}
\vspace{-3pt}
\label{tab:result-cola-plus}
\centering
\begin{tabular}{l|r|r}
\hline
Method&  Average &  Improvement \\
\hline
\hline
COLA'          &    60.9\% &                     \\
\hline
COLA'+Gaussian  &    62.7\% &           COLA' +1.8 \\
COLA'+Mixup     &    66.7\% &           COLA' +5.8 \\
COLA'+Mixup+RRC &    68.2\% &           COLA' +7.3 \\
\hline
\hline
BYOL-A (Mixup+RRC)  &  \textbf{72.3\%} &  \\
\hline
BYOL-A (Mixup only) &    63.9\% &  BYOL-A -8.4  \\
BYOL-A (RRC only) &    68.4\% &  BYOL-A -3.9  \\
\hline
\end{tabular}
\end{table}

We assessed the effectiveness of data augmentation blocks in the BYOL-A augmentation module by adding augmentation blocks to COLA', named COLA'+; the original COLA does not use data augmentations.
Table \ref{tab:result-cola-plus} shows that improvement of COLA'+Mixup is $5.8$, larger than Gaussian-noise's $1.8$. Mixup is also effective for making useful contrast with two differently random cropped input segments.

However, RRC was not as effective as when it was used with BYOL-A. Adding RRC to COLA'+Mixup improves performance to $68.2 - 66.7 = 1.5$, which is less performance gain compared to when RRC is applied to BYOL-A (Mixup only), where the improvement is $72.3 - 63.9 = 8.4$. The explanation for this is that additional random resizing and cropping to segments that have already been randomly cropped is less effective; and the RRC-only setting of BYOL-A performs better.

In summary, the BYOL-A augmentation module is also effective with COLA'+, but more effective with BYOL, a single segment input setting.

\subsection{Experiment for pretraining on FSD50K} \label{appendix:fsd50k}

\begin{table}[tb!]
\vspace{-10pt}
\caption{Average performance of BYOL-A with pretraining datasets}
\vspace{-5pt}
\label{tab:perf-pretraining-fsd50k}
\centering
\begin{tabular}{ll|r|r}
\hline
Pretraining dataset & Size &  Average & Difference \\
\hline
\hline
AudioSet (1/10 subset) & 210K &    \textbf{72.3\%}&  \\
\hline
FSD50K         & 40K  &    70.1\% & AudioSet -2.2 \\
\hline
\end{tabular}
\vspace{-13pt}
\end{table}

In addition to AudioSet, we conducted pretraining on the FSD50K\cite{fonseca2020fsd50k} dataset and compared performance with pretraining on AudioSet.

All the training hyper-parameters and the setup were the same as in the former experiments, except the number of pretraining epochs was set to 500 on FSD50K, so that the total number of data samples consumed during training would be closer to the experiments on the AudioSet 1/10 subset.
In addition, we used the development subset of the FSD50K, which has 40,966 samples, five times less than the AudioSet 1/10 subset (210,315 samples in total).

As shown in Table \ref{tab:perf-pretraining-fsd50k}, the average performance for FSD50K is 70.1\%; the difference from AudioSet is -2.2. We think this degradation is caused by the smaller data size. Though the performance is lower than that for AudioSet pretraining, it still outperforms conventional methods shown in the Table \ref{tab:result-dstasks}.

\bibliographystyle{IEEEtran}
\bibliography{refs}

\end{document}